\documentclass[conference,10pt]{IEEEtran}

\usepackage{graphicx}
\usepackage{amsmath}
\usepackage{amssymb}
\usepackage{booktabs}
\usepackage{array}
\usepackage{xcolor}
\usepackage[hidelinks]{hyperref}
\usepackage{url}

\newcommand{\tabhead}[1]{\textbf{#1}}

\begin{document}

\title{Lesion-Gated Hybrid Synthesis for Virtual\\
Contrast-Enhanced Breast MRI:\\
A MAMA-SYNTH Challenge Solution}

\author{\IEEEauthorblockN{Shohei Yoshimoto}
\IEEEauthorblockA{Department of Radiology, The Jikei University School of Medicine\\
Tokyo, Japan}}

\maketitle

\begin{abstract}
\textit{Purpose:} Contrast-enhanced breast MRI depends on intravenous
gadolinium-based contrast agents, motivating methods that synthesise post-contrast
appearance from pre-contrast images alone. The MAMA-SYNTH Challenge (MICCAI 2026
Deep-Breath Workshop) evaluates such synthesis across four metric groups---image
fidelity, tumour region of interest, downstream classification, and downstream
segmentation---ranked by the average of the four group ranks, so that optimising a
single objective is insufficient.
\textit{Materials and Methods:} We developed a lesion-gated hybrid synthesis
pipeline. A lesion probability map, estimated from the pre-contrast slice alone by an
ensemble of pre-contrast-only segmentation networks, spatially coordinates a
tumour-focused regression pathway and a background-focused Pix2PixHD synthesis
pathway, followed by a region-dependent calibration of the predicted enhancement.
Training used the public MAMA-MIA collection with a patient-level split. Inference
consumes a single pre-contrast 2D slice, with no mask, no post-contrast image, and no
auxiliary metadata beyond image geometry.
\textit{Results:} On internal validation ($n=120$ patients) the method reached MSE
0.523, LPIPS 0.185, and tumour-region SSIM 0.495. It was submitted to the hidden
external 300-case test cohort as a self-contained inference container. The method
ranked sixth in the official MAMA-SYNTH Challenge leaderboard.
\textit{Conclusion:} A test-compatible lesion probability map derived from the
pre-contrast image can coordinate complementary tumour-focused regression and
background-focused perceptual synthesis, enabling balanced virtual contrast
enhancement under a multi-metric challenge setting.
\end{abstract}


\section{Introduction}

Dynamic contrast-enhanced (DCE) MRI is central to breast cancer detection, staging,
and treatment-response assessment, but it requires intravenous gadolinium-based
contrast agents, which carry acute-reaction risk, deposition and retention concerns,
cost, and workflow burden~\cite{ref:gbca}. Synthesising the post-contrast appearance
from pre-contrast acquisitions---virtual contrast enhancement---has therefore
attracted sustained interest, with reader studies reporting that a substantial
fraction of synthesised examinations can reach diagnostic
quality~\cite{ref:mullerfranzes,ref:chung}.

What makes the problem difficult is not the regression itself but what a clinically
meaningful synthetic image must simultaneously satisfy. It must be numerically
faithful to the true post-contrast image; it must look like a real MR image rather
than a smoothed estimate; and it must be correct \emph{where it matters}, inside the
lesion, so that downstream automated readers behave as they would on a real
acquisition. In our experiments these requirements did not co-optimise. A residual
U-Net reduced mean squared error by approximately 55\% relative to the unprocessed
input, yet produced \emph{worse} perceptual similarity than the input itself;
conversely, an adversarial generator achieved the best perceptual score of any single
model we trained while degrading pixel error by 29\% and tumour-region structure by
24\%.

The MAMA-SYNTH Challenge~\cite{ref:mamasynth} makes this tension the explicit object
of evaluation, scoring eight metrics in four groups and ranking by the mean of the
four group ranks. A method that dominates one group while failing another is penalised
by construction. The challenge further constrains the solution space in two ways that
shaped every decision we made: inference receives a \textbf{single pre-contrast 2D
slice} with no mask and no auxiliary metadata beyond image geometry, and the test
phase permits \textbf{exactly one submission}.

This paper is a challenge solution report rather than a controlled methodological
study. It describes the engineering design we arrived at under those constraints, the
internal evidence on which each design decision was made, and the outcome on the
hidden test cohort. Our solution is a modular pipeline with an explicit division of
labour: (i) a \emph{lesion localization} ensemble provides spatial guidance, computed
from the pre-contrast image alone; (ii) a \emph{regression pathway} carries anatomical
fidelity and lesion enhancement; (iii) a \emph{Pix2PixHD pathway} contributes
perceptual realism and background texture; and (iv) a \emph{gated composition}
controls the trade-off between them region by region.

\textbf{Contributions.} (i) We describe a virtual contrast enhancement solution in
which a test-compatible lesion probability map coordinates complementary
tumour-focused regression and background-focused perceptual synthesis, followed by a
region-dependent calibration; the method ranked sixth in the official MAMA-SYNTH
Challenge leaderboard. (ii) We report the fidelity/perception/conspicuity trade-off we
observed on a common internal validation split, on which none of the model families we
evaluated provided the desired balance alone. (iii) We report an internal ablation of
how lesion-localization quality and gating strategy each affected synthesis
performance, including a ground-truth-mask oracle that bounds the remaining headroom,
and the design lessons we drew from it.

\section{Challenge and Data}

\textbf{Task and inference contract.} The task is to synthesise the peak-enhancement
post-contrast 2D slice from the corresponding pre-contrast T1-weighted fat-saturated
slice. Submissions are algorithm containers. At inference the container receives one
2D float32 image in z-score space and must emit one image of identical shape with
spacing, origin, and direction preserved. No segmentation mask, no post-contrast
image, no neighbouring slices, no 3D volume, and no auxiliary clinical metadata are
provided at test time; image geometry is the only side information available, and it
is copied through to the output. The test phase accepts a single submission, and a
technical report is mandatory.

\textbf{Cohort and preprocessing.} Training data is the public MAMA-MIA
collection~\cite{ref:mamamia}: 1{,}506 patients drawn from four TCIA collections
(DUKE 291, ISPY1 171, ISPY2 980, NACT 64) across more than 25 US centres, with GE
(966), Siemens (411), and Philips (129) scanners, all cases biopsy-confirmed. We
applied the official challenge preprocessing~\cite{ref:mamasynthcode} without
modification: the peak phase is the DCE phase with maximum mean intra-tumour signal in
3D, the slice is the one with the largest tumour area along the through-plane axis, and
both pre and peak images are normalised with dataset-level statistics (mean 107.41194,
std 219.96181) and written as float32. All 1{,}506 cases processed successfully.
In-plane sizes are heterogeneous---seven distinct sizes, including $312\times312$,
which is not a multiple of 32---so all models operate on reflect-padded inputs and all
evaluation is performed at native resolution.

\textbf{Split.} From the MAMA-MIA training portion we carved a \textbf{patient-level}
split of 1{,}080 training and 120 internal validation patients; the dataset-internal
held-out set of 306 patients was excluded from training throughout. The five-fold
cross-validation used for lesion localization was performed strictly within the
1{,}080 training patients.

\textbf{Hidden test cohort.} The challenge test set comprises 300 cases from two
external institutions in the Netherlands and Argentina---Radboud UMC (200 cases,
$416\times416$, Siemens 3T) and Instituto Alexander Fleming (100 cases,
$512\times512$, GE 1.5T)---acquired outside the training institutions and geographic
distribution, with different acquisition settings and population characteristics.

\textbf{Evaluation.} Eight metrics in four groups: image fidelity (MSE $\downarrow$,
LPIPS $\downarrow$~\cite{ref:lpips}), tumour region of interest (tumour SSIM
$\uparrow$~\cite{ref:ssim}, Fr\'echet Radiomic Distance
$\downarrow$~\cite{ref:frd}), classification (AUROC pre vs.\ post $\uparrow$, AUROC
tumour vs.\ non-tumour ROI $\uparrow$), and segmentation (Dice $\uparrow$, HD95
$\downarrow$). Per-case values are averaged and converted to a rank; the two ranks
within a group are averaged; the four group ranks are averaged; lowest wins, ties
broken in the order ROI, classification, segmentation, image. The classification and
segmentation metrics are produced by \textbf{organizer-fixed pre-trained
evaluators}---a single-fold 2D nnU-Net~\cite{ref:nnunet} and a radiomics classifier
ensemble---applied to the \emph{synthesised} image. They measure a downstream property
of our output, not our own segmentation or classification performance.

\textbf{External data policy.} Only public data and public pretrained weights are
permitted, and all must be declared. We used no dataset other than MAMA-MIA;
pretrained weights are itemised in Section~\ref{sec:data-masks}.

\begin{table}[t]
\caption{Challenge and data setup.}
\label{tab:setup}
\centering
\footnotesize
\renewcommand{\arraystretch}{1.12}
\begin{tabular}{@{}p{0.28\columnwidth}p{0.63\columnwidth}@{}}
\toprule
\tabhead{Item} & \tabhead{Value} \\
\midrule
Task & Pre-contrast T1 fat-saturated 2D slice $\rightarrow$ peak-enhancement post-contrast 2D slice \\
Inference input & One 2D float32 z-score image; no mask, no auxiliary metadata beyond image geometry, no 3D context \\
Inference output & One image, identical shape; spacing, origin, direction preserved \\
Submission & Algorithm container; test phase limited to one submission \\
Cohort & MAMA-MIA, 1{,}506 patients; DUKE 291 / ISPY1 171 / ISPY2 980 / NACT 64; $>$25 US centres \\
Scanners & GE 966 / Siemens 411 / Philips 129 \\
Preprocessing & Official pipeline, unmodified; peak phase by maximum mean intra-tumour signal in 3D; largest-tumour-area slice; dataset z-score (107.41194, 219.96181) \\
Native sizes & 7 distinct; $256^2$ (983), $512^2$ (193), $384^2$ (140), $448^2$ (99), $320^2$ (70), $312^2$ (19), $128^2$ (2) \\
Split & 1{,}080 train / 120 internal validation, patient-level; 306 dataset-internal cases excluded \\
Hidden test & 300 cases at two external institutions: Radboud UMC, Netherlands 200 ($416^2$, Siemens 3T); Instituto Alexander Fleming, Argentina 100 ($512^2$, GE 1.5T) \\
Metric groups & Image (MSE $\downarrow$, LPIPS $\downarrow$); ROI (tumour SSIM $\uparrow$, FRD $\downarrow$); Classification (2 $\times$ AUROC $\uparrow$); Segmentation (Dice $\uparrow$, HD95 $\downarrow$) \\
Ranking & Per-metric rank $\rightarrow$ mean within group $\rightarrow$ mean of four group ranks; lowest wins; ties broken ROI, CLF, SEG, Image \\
External data & Public data and public pretrained weights only; all must be declared \\
\bottomrule
\end{tabular}
\end{table}

\section{Method}

The submitted method is a \emph{lesion-gated hybrid synthesis} pipeline (internal
identifier \texttt{primary\_calibrated\_v2}): five pre-contrast-only lesion
localization networks, three regression networks, and one adversarial generator,
composed by two gates derived from a single spatial prior and closed by a
region-dependent calibration.

\begin{figure*}[t]
\centering
\includegraphics[width=0.976\textwidth]{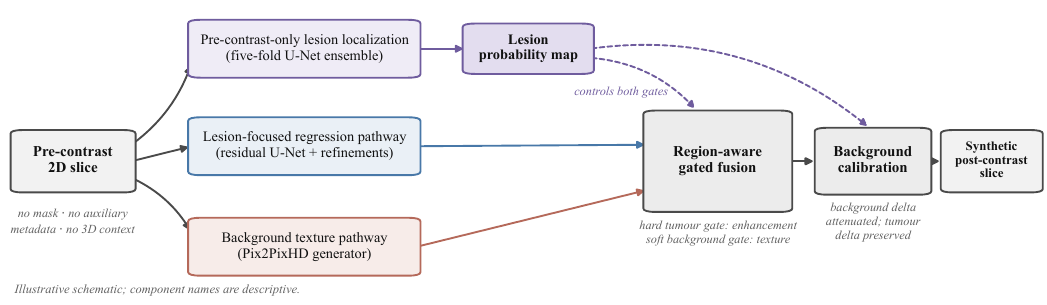}
\caption{Illustrative schematic of the lesion-gated hybrid synthesis pipeline. A lesion
probability map is estimated from the pre-contrast slice alone and drives two gates: a
hard dilated gate through which the lesion-focused regression pathway supplies
enhancement, and a complementary soft gate through which the Pix2PixHD-based pathway
supplies background texture. A final region-dependent calibration attenuates the
predicted background enhancement while leaving tumour enhancement unchanged. Inference
requires no mask, no auxiliary metadata beyond image geometry, and no 3D context.
Component names in the figure are descriptive rather than architectural identifiers;
exact configurations are given in Section~III.}
\label{fig:pipeline}
\end{figure*}

\subsection{Design principle}

Three design choices shaped the pipeline. First, \textbf{anatomy is carried, not
re-synthesised}: every regression component predicts an enhancement \emph{residual}
added to the input, so the pre-contrast image is explicitly retained as the residual
reference, reducing the need to re-synthesise anatomy from scratch. Second,
\textbf{components are allowed to specialise}: rather than asking one network to be
simultaneously accurate, realistic, and tumour-correct, we train components that are
each permitted to be unbalanced. Third, \textbf{composition requires a spatial prior
that survives deployment}: the decision of which component acts where must be
computable from the pre-contrast image alone.

\subsection{Lesion probability from the pre-contrast image alone}
\label{sec:lp}

An ensemble of five U-Nets~\cite{ref:unet} (base width 32, depth 4, group
normalisation) is trained as a binary pre-contrast-only segmentation task with a Dice
plus focal binary cross-entropy objective, using five-fold cross-validation within the
1{,}080 training patients. Out-of-fold predictions provide lesion probability maps for
training cases; the fold ensemble provides them for validation cases. Per-fold
held-out Dice ranged from 0.633 to 0.679, and the five-fold ensemble reached Dice
0.702 on the internal validation split under our own implementation.

The submitted container used \textbf{mean-logit aggregation across the five
segmentation folds}:
\begin{equation}
\mathrm{LP} = \sigma\!\left(\tfrac{1}{5}\textstyle\sum_{i=1}^{5} f_i(x_{\mathrm{pre}})\right).
\end{equation}
The choice of aggregation rule matters at deployment. At deployment, all five models
are evaluated on each unseen test case, whereas training-case out-of-fold maps are
produced by held-out models. A maximum-style aggregation therefore tends to produce
broader probability maps than the out-of-fold maps used during development; the deployed
mean-logit gate covers approximately 85.7\% of the out-of-fold gate area at threshold
0.20. This component is what makes the entire
composition legal at test time: no ground-truth mask is required to decide where each
branch acts.

\subsection{Synthesis branches}

\textbf{Base residual U-Net.} A residual U-Net (base width 32, depth 4) predicting the
enhancement residual, trained with $L_1$ plus gradient $L_1$ (0.1), ROI $L_1$ (0.2)
and a gentle ROI SSIM term (0.05). It supplies anatomical fidelity and is the base
prediction for both refinement branches.

\textbf{Contrast-preserving refinement.} A two-channel refiner taking the base
prediction and the pre-contrast image, trained with additional asymmetric-contrast,
background-over-enhancement, tumour-under-enhancement, ROI texture and
lesion-probability-weighted terms. It improves tumour--background contrast at
essentially no cost in pixel error. The lesion probability map enters this component
only as a \emph{loss weight} during training; it is not a network input.

\textbf{Tumour-enhancement refinement.} A refiner of identical architecture,
initialised from the contrast-preserving refiner and fine-tuned at a low learning rate
with markedly stronger tumour-under and asymmetric-contrast penalties. It is
intentionally over-enhancing, and is therefore admitted only inside the lesion gate and
only in the positive direction.

\textbf{Pix2PixHD-based texture synthesis.} A Pix2PixHD global
generator~\cite{ref:pix2pixhd} (64 base filters, four downsamplings, nine residual
blocks) trained adversarially with LSGAN, $L_1$, VGG-perceptual~\cite{ref:vgg},
feature-matching, ROI $L_1$, and asymmetric tumour/background terms. It supplies the
high-frequency texture that regression losses average away.

The generator is the only component trained in a per-image $[-1,1]$ space while all
others operate in the dataset z-score space. The deployed container therefore converts
across this boundary explicitly, using per-image $r_{\min}, r_{\max}$ and the dataset
constants $\mu_{\mathrm{ds}}=107.41194$, $\sigma_{\mathrm{ds}}=219.96181$:
\begin{align}
x_{\mathrm{raw}} &= \max\!\left(0,\; x_z \sigma_{\mathrm{ds}} + \mu_{\mathrm{ds}}\right),\\
x_{\mathrm{p2p}} &= \mathrm{clip}\!\left(2\tfrac{x_{\mathrm{raw}} - r_{\min}}{r_{\max} - r_{\min}} - 1,\, -1,\, 1\right),
\end{align}
with the inverse mapping applied to the generator output before fusion. We note this
explicitly because the mismatch was present in an intermediate build and was caught by
train/inference tensor parity checking before the container was packaged;
component-wise normalisation parity is, in our experience, a first-order
reproducibility risk in multi-component pipelines.

\subsection{Lesion-guided gated fusion}

Two gates are derived from the same lesion probability map, and their separation is the
core of the method. A \textbf{hard} gate admits tumour enhancement,
\begin{equation}
g_{\mathrm{boost}} = \mathrm{dilate}\!\left(\mathrm{LP} \ge 0.20,\ 8 \text{ iterations}\right),
\end{equation}
the dilation providing a margin so the boost covers the lesion rim rather than only its
confident core. A \textbf{soft} gate controls texture,
\begin{equation}
g_{\mathrm{blend}} = \sigma\!\left(\frac{\mathrm{LP} - 0.20}{0.10}\right),\quad
g_{\mathrm{blend}} = 0 \ \text{where} \ \mathrm{LP} < 0.01 .
\end{equation}
Writing $b$ for the base prediction, $c$ for the contrast-preserving refinement, $t$
for the tumour-enhancement refinement and $p$ for the generator output,
\begin{align}
t^{+} &= c + \max(0,\, t - c),\\
u &= b\,(1 - g_{\mathrm{boost}}) + \max(b,\, t^{+})\, g_{\mathrm{boost}},\\
\hat{y} &= u + 0.35\,(1 - g_{\mathrm{blend}})\,(p - u).
\end{align}
In words: \textbf{inside the lesion, enhancement is decided by the regression pathway;
outside it, the generator contributes texture.} The one-sided $\max$ operators ensure
the tumour-enhancement branch can only add signal where the gate is open, never
subtract it.

\subsection{Background delta calibration}

The fused prediction retains a systematic tendency to over-enhance background tissue.
The final stage scales the predicted enhancement in a region-dependent manner,
\begin{align}
s &= 0.90\,(1 - g_{\mathrm{boost}}) + 1.00\, g_{\mathrm{boost}},\\
y &= x_{\mathrm{pre}} + s \odot (\hat{y} - x_{\mathrm{pre}}).
\end{align}
A conservative calibration setting was selected to suppress background
over-enhancement while preserving tumour enhancement; it was not the numerically best
configuration in the internal sweep (Section~\ref{sec:abl-calib}).

\subsection{Training data, mask usage, and pretrained weights}
\label{sec:data-masks}

\textbf{Raw-derived neighbour augmentation.} To enlarge the adversarial training set
without violating the single-slice inference contract, we extracted additional 2D
pairs from the raw 3D volumes of the \textbf{training patients only}, at slice offsets
$z = 0, \pm 1, \pm 2$ around the officially selected slice, yielding \textbf{5{,}386
pairs}. A weighted sampler (1.0 / 0.8 / 0.5 by offset) gives 3{,}881 effective samples
per epoch. These pairs were used \textbf{only} to fine-tune the Pix2PixHD generator;
internal validation retained the original centre slices unmodified, and inference
remains single-slice throughout. Inter-slice information therefore enters this work
purely as training-time augmentation.

\textbf{Mask usage.} Expert lesion masks were used as segmentation targets,
loss-region weights, and internal evaluation regions of interest, and were never used
as model inputs. The submitted container accepts no mask input.

\textbf{Pretrained weights.} An ImageNet-pretrained VGG19 network~\cite{ref:vgg} was
used for the perceptual loss term during Pix2PixHD training only, and an AlexNet-based
LPIPS network was used as a training and validation monitor only. Neither is present at
inference: the \textbf{submitted container contains no external pretrained model},
only the nine checkpoints trained in this work (five localization folds, three
refinement/base networks, one generator). No nnU-Net pretrained weights distributed
with the dataset were downloaded or used.

\textbf{Deployment.} The container is Linux/amd64, runs as a non-root user, has no
network access, and bakes all weights at build time (6.41\,GB compressed). It
validates output shape, dtype, and finiteness, and copies image geometry from the
input.

\section{Experimental Setup}

\textbf{Training.} All U-Net-family components use base width 32, depth 4, group
normalisation with 8 groups, and reflect padding to a multiple of 32. Learning rates
were $3\times10^{-4}$ for the base residual network (up to 80 epochs) and for each
localization fold (up to 60 epochs), $1\times10^{-4}$ for the contrast-preserving
refiner, and $1\times10^{-5}$ for the tumour-enhancement refiner initialised from it.
The generator (182.4\,M parameters, with a 5.5\,M-parameter multi-scale discriminator)
was fine-tuned on the neighbour-augmented set at $2\times10^{-5}$ and $5\times10^{-5}$
for generator and discriminator, best checkpoint at epoch 2, early stopping at epoch
12. Augmentation was horizontal flipping only. Training used a single RTX 4500 Ada GPU.

\textbf{Evaluation tiers.} We distinguish four strictly separate tiers of evidence and
never compare across them: \emph{internal validation} (120 patients, patient-level
held out, our own metric implementations---used for development and all ablations);
\emph{challenge debug phase} (5 fixed cases, official metrics---candidate selection
only); \emph{challenge validation phase} (official metrics---a design
diagnostic); and \emph{challenge test phase} (hidden, 300 cases, official
metrics---the final outcome).

\textbf{Internal metric caveats, stated once and applied throughout.} Our internal
implementations of LPIPS and tumour-region SSIM are not the challenge evaluator's
implementations---the official ROI SSIM computes a full-image SSIM map and averages it
inside the mask with a fixed data range---so internal values are comparable only within
our own harness and are not predictions of official values. Two internal evaluation
runs with different metric implementations exist in this work; every table states
which run produced it, and no table mixes them. In addition, tumour-under,
background-over, and contrast-error are in-house diagnostics defined for error
analysis, not challenge metrics.

\textbf{Internal validation is a development set, not a holdout.} The same 120 patients
were used to select architectures, loss weights, the lesion probability source, fusion
weights, gate thresholds, dilation radii, post-processing, and calibration
constants---on the order of 157 gate and calibration variants alone. Internal numbers
should therefore be read as optimistic by an unquantified margin. This is the principal
reason the final candidate decision was made on official same-cohort evidence rather
than on internal ranking (Section~\ref{sec:selection}).

\section{Results and Ablations}

\subsection{Challenge result}

The final method was evaluated on the hidden external 300-case test cohort and
\textbf{ranked sixth in the official MAMA-SYNTH Challenge leaderboard}.

Per-metric test results and the composite score had not been publicly released at the
time of writing, and we therefore do not estimate them. No debug-phase,
validation-phase, or internal number in this paper should be read as a substitute for
them.

On internal validation ($n=120$ patients, our own metric implementation) the submitted
configuration reached MSE 0.523, LPIPS 0.185, and tumour-region SSIM 0.495.

\begin{table*}[t]
\caption{Results by evaluation tier. Tiers are never compared with one another.
Block (b) lists raw official metrics on a fixed five-case cohort and is
candidate-selection evidence only; no rank or composite score from that phase is
reported. Block (c) describes an earlier candidate, not the submitted method.}
\label{tab:results}
\centering
\footnotesize
\renewcommand{\arraystretch}{1.15}
\setlength{\tabcolsep}{4.2pt}
\begin{tabular}{@{}lccccccccc@{}}
\toprule
\tabhead{Configuration} & \tabhead{MSE $\downarrow$} & \tabhead{LPIPS $\downarrow$} &
\tabhead{ROI SSIM $\uparrow$} & \tabhead{FRD $\downarrow$} & \tabhead{AUROC$_{\mathrm{C}}\uparrow$} &
\tabhead{AUROC$_{\mathrm{T}}\uparrow$} & \tabhead{Dice $\uparrow$} & \tabhead{HD95 $\downarrow$} \\
\midrule
\multicolumn{9}{@{}l}{\emph{(a) Internal validation, $n=120$, our own metric implementation}}\\
\quad Submitted method & 0.523 & 0.185 & 0.495 & --- & --- & --- & --- & --- \\
\addlinespace[2pt]
\multicolumn{9}{@{}l}{\emph{(b) Challenge debug phase, fixed 5-case cohort, official metrics --- candidate selection only}}\\
\quad Earlier ensemble & 0.716 & 0.207 & 0.367 & 28.39 & 0.88 & 0.42 & 0.348 & 245.96 \\
\quad\quad + perceptual post-processing & 0.725 & 0.166 & 0.366 & 29.03 & 0.88 & 0.38 & 0.345 & 246.42 \\
\quad Submitted method & 0.569 & 0.159 & 0.360 & 27.88 & 0.84 & 0.62 & 0.347 & 243.42 \\
\addlinespace[2pt]
\multicolumn{9}{@{}l}{\emph{(c) Challenge validation phase, official metrics --- earlier candidate, snapshot 2026-06-24}}\\
\quad Earlier ensemble & 0.794 & 0.137 & 0.488 & 26.17 & 0.805 & 0.655 & 0.345 & 246.52 \\
\quad\quad group mean ranks & \multicolumn{2}{c}{Image 42.0} & \multicolumn{2}{c}{ROI 22.5} &
\multicolumn{2}{c}{Classification 28.0} & \multicolumn{2}{c}{Segmentation 30.5} \\
\addlinespace[2pt]
\multicolumn{9}{@{}l}{\emph{(d) Challenge test phase, hidden cohort, $n=300$, official metrics}}\\
\quad Submitted method --- \textbf{official final rank: 6th} & \multicolumn{8}{c}{\emph{per-metric results not publicly released at the time of writing}} \\
\bottomrule
\end{tabular}
\end{table*}

\subsection{How the final design was selected}
\label{sec:selection}

Two pieces of official feedback obtained before the single test submission shaped the
final method. Both are candidate-selection evidence; neither is a proxy for test-phase
performance.

\textbf{The image group was the binding constraint.} An earlier region-ensemble
candidate scored in the challenge validation phase produced per-group mean ranks of
42.0 (image), 22.5 (ROI), 28.0 (classification), and 30.5 (segmentation) at the
snapshot taken on its submission date. Tumour-region performance was already
competitive while image fidelity and perceptual similarity lagged. Because the
composite score is a rank average against a moving field, its absolute rank drifted as
the field grew, and we therefore treat only the per-group structure as informative.

\textbf{Perceptual post-processing failed a multi-group test.} On a fixed five-case
debug cohort ($n=5$), applying a high-frequency merge and unsharp mask to that
candidate improved official LPIPS by 19.9\% but degraded the tumour-versus-non-tumour
AUROC by 9.5\% and slightly worsened MSE. Unmasked sharpening raises perceptual scores
everywhere, \emph{including inside the lesion}, where it perturbs the radiomic
signature the organizer-fixed classification evaluator reads. This is the direct origin
of the decision to restrict generative texture to the background.

\textbf{Candidate selection under a one-shot budget.} On the same fixed five-case debug
cohort ($n=5$), the submitted method improved raw MSE by 20.5\%, LPIPS by 22.9\%,
Fr\'echet Radiomic Distance by 1.8\%, and HD95 by 1.0\% relative to the earlier
ensemble, with Dice essentially tied and tumour-region SSIM 1.9\% lower. With $n=5$---two
of the five cases have Dice 0---these values support a selection decision and nothing
more, and we do not interpret the AUROCs at this cohort size. Their value was
methodological: our internal proxies had ranked these two candidates in the
\emph{opposite} order, so only a same-cohort official measurement resolved the choice
before the irreversible submission.

\subsection{Ablations}

All ablations are internal ($n=120$, our own metrics). Table~\ref{tab:ablation} states
which evaluation run produced each block.

\textbf{(a) The regression backbone preserves anatomy but degrades realism.} Relative
to the unprocessed input, the base residual U-Net reduces MSE from 1.284 to 0.571 and
raises the internal tumour-region structural score from 0.195 to 0.560---but
\emph{worsens} LPIPS from 0.149 to 0.202. Contrast-preserving refinement then improves
in-house contrast error by 8.2\% and tumour-under-enhancement by 10.8\% at essentially
unchanged MSE. The region ensembles built from these components add little beyond them.
Regression alone did not reach the image group, because the loss that buys pixel
accuracy is the loss that removes texture.

\textbf{(b) The generator is the best perceptual model and the weakest tumour model.}
Trained standalone, the Pix2PixHD generator achieves LPIPS 0.144---the best of any
single model in this work, beating the base U-Net in 116 of 120 cases---at MSE 0.737
($+29\%$) and internal tumour-region SSIM 0.424 ($-24\%$). It also carries a structural
limitation: its bounded output saturates against the input's intensity range,
truncating strong enhancement in roughly 0.29\% of pixels, concentrated precisely at
tumour peaks. The generator is therefore admitted only through the background-weighted
blend and never as the tumour predictor.

\begin{figure*}[t]
\centering
\includegraphics[width=0.96\textwidth]{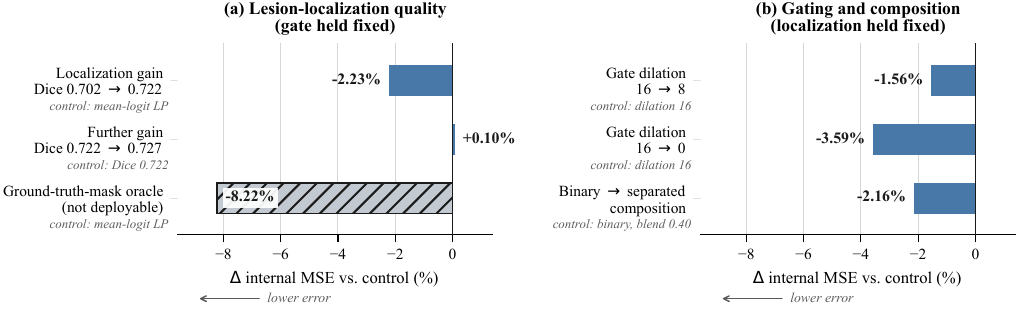}
\caption{Effects of lesion-localization quality and gating strategy on internal
synthesis performance. Internal validation, $n=120$, our own metric implementation, a
single base pipeline. Each bar is a controlled one-factor comparison against the named
control; negative values indicate lower error. The ground-truth-mask oracle is not
deployable. Dice values are joined from a separate segmentation evaluation, and the
mean-logit Dice ($\approx 0.702$) is approximate.}
\label{fig:gate}
\end{figure*}

\textbf{(c) Localization quality and gating strategy both affected synthesis
performance.} Fig.~\ref{fig:gate} presents this analysis. All values come from a single
sweep series on one base pipeline; each is a controlled one-factor comparison against
the named control.

Improving the lesion-probability source from the deployed mean-logit aggregation (Dice
$\approx 0.702$) to a raw-3D-pretrained model (Dice 0.722), with the gate held fixed,
improved internal MSE by \textbf{2.23\%}. A further increase to Dice 0.727, obtained by
combining two such models with surface suppression, changed MSE by \textbf{$+0.10\%$}
---that is, localization gains had flattened by this point. With the
lesion-probability source held fixed, reducing the gate dilation from 16 to 0 improved
MSE by \textbf{3.59\%}, and replacing the single binary gate with the separated
hard-boost / soft-blend composition at a blend weight of 0.35 improved MSE by
\textbf{2.16\%} while reducing per-case MSE regressions from 38 to 26 of 120. Replacing
the predicted map with the ground-truth mask---an oracle that is not deployable---improved
MSE by \textbf{8.22\%}; a \emph{dilated} ground-truth mask recovered only 1.47\%,
indicating that the oracle benefit derives from the precision of the region rather than
from the mere availability of a mask.

Summarising, within the explored range both lesion-localization quality and the
downstream gating and composition strategy affected synthesis performance. Localization
gains showed diminishing returns at higher Dice, while the oracle analysis indicated
substantial remaining headroom. Each localization effect replicated at three
independent dilation settings, so it is not an artifact of one gate configuration.

\textbf{(d) Neighbour augmentation improved tumour fidelity without collateral
damage.} Fine-tuning the generator on the 5{,}386 neighbour-derived pairs reduced its
standalone tumour-under-enhancement by 27.6\% and in-house contrast error by 11.5\%, at
unchanged MSE and LPIPS, while increasing background over-enhancement by 10.2\%. Inside
the full pipeline most of this is absorbed: LPIPS improves 1.05\%,
tumour-under-enhancement improves 1.07\%, MSE moves 0.34\% in the worse direction, and
the background penalty shrinks to 2.8\% because the calibration stage removes most of
it. Of the cases whose MSE worsened by more than 5\%, all ten were in the low-error
regime (maximum absolute MSE 0.034) and none were high-error cases. This checkpoint
change is precisely the difference between the submitted method and its immediate
predecessor.

\textbf{(e) Gating the calibration mattered more than its magnitude.}
\label{sec:abl-calib}
In a 45-variant sweep spanning uncalibrated, globally scaled,
lesion-probability-linear, high-error-case-only, and gated families, every gated
variant outperformed every ungated one on the sweep's internal composite. A
\emph{global} 0.90 scaling improves MSE but costs 22.8\% in tumour-under-enhancement---it
buys the image group by damaging the ROI group---whereas the gated form leaves tumour
enhancement untouched. A conservative calibration setting (background 0.90, tumour
1.00) was selected to suppress background over-enhancement while preserving tumour
enhancement; it was not the numerically best configuration in the internal sweep. A
slightly more aggressive variant scored better on the internal composite but raises
tumour enhancement above unity, an extrapolation our internal evaluation on 120
development cases cannot validate.

\begin{table*}[t]
\caption{Internal ablations ($n=120$, our own metric implementation). Blocks come from
different evaluation runs and their absolute values are \textbf{not} comparable across
blocks; only within-block differences are meaningful. Contrast error (CE),
background-over (BG) and tumour-under (TU) are in-house diagnostics, not challenge
metrics. The submitted method's headline internal values appear in
Table~\ref{tab:results}(a), a third evaluation run, and are not differenced against
either block.}
\label{tab:ablation}
\centering
\footnotesize
\renewcommand{\arraystretch}{1.12}
\begin{tabular}{@{}llcccccc@{}}
\toprule
& \tabhead{Configuration} & \tabhead{MSE $\downarrow$} & \tabhead{LPIPS $\downarrow$} &
\tabhead{ROI SSIM$^{\dagger}\uparrow$} & \tabhead{CE $\downarrow$} & \tabhead{BG $\downarrow$} & \tabhead{TU $\downarrow$} \\
\midrule
\multicolumn{8}{@{}l}{\emph{Block A --- component comparison (2026-07-01 evaluation run)}}\\
& Unprocessed input & 1.2837 & 0.1489 & 0.1952 & 3.0428 & 0.0264 & 3.3592 \\
& Base residual U-Net & 0.5710 & 0.2021 & 0.5599 & 0.8615 & 0.1556 & 0.8905 \\
& Contrast-preserving refinement & 0.5709 & 0.2084 & 0.5665 & 0.7910 & 0.1333 & 0.7940 \\
& Region-separated branch & 0.5409 & 0.2048 & 0.5639 & 0.8222 & 0.1183 & 0.8582 \\
& Region ensemble & 0.5466 & 0.2051 & 0.5638 & 0.8115 & 0.1195 & 0.8285 \\
& Region ensemble, improved LP & 0.5436 & 0.2050 & 0.5637 & 0.8068 & 0.1195 & 0.8223 \\
& Pix2PixHD generator, standalone & 0.7370 & 0.1440 & 0.4240 & --- & --- & --- \\
\addlinespace[3pt]
\multicolumn{8}{@{}l}{\emph{Block B --- gating, composition and calibration (2026-07-03/04 sweep run)}}\\
& \multicolumn{7}{@{}l}{\emph{Lesion-probability source, gate fixed (threshold 0.20, dilation 16)}}\\
& \quad mean-logit LP (deployed), Dice $\approx$0.702 & 0.5595 & --- & --- & --- & --- & --- \\
& \quad raw-3D pretrained, Dice 0.722 & 0.5470 & --- & --- & --- & --- & --- \\
& \quad two-model combination, Dice 0.727 & 0.5476 & --- & --- & --- & --- & --- \\
& \quad ground-truth-mask oracle (not deployable) & 0.5135 & --- & --- & --- & --- & --- \\
& \multicolumn{7}{@{}l}{\emph{Gate dilation, lesion-probability source fixed}}\\
& \quad dilation 16 & 0.5476 & 0.1636 & 0.5722 & 0.7856 & 0.1296 & 0.7388 \\
& \quad dilation 8 & 0.5390 & 0.1621 & 0.5702 & 0.7910 & 0.1266 & 0.7488 \\
& \quad dilation 0 & 0.5279 & 0.1605 & 0.5676 & 0.8012 & 0.1221 & 0.7852 \\
& \multicolumn{7}{@{}l}{\emph{Composition, source and dilation fixed}}\\
& \quad binary gate, blend 0.40 & 0.5390 & 0.1621 & 0.5702 & 0.7910 & 0.1266 & 0.7488 \\
& \quad separated gate, blend 0.35 & 0.5274 & 0.1668 & 0.5710 & 0.7926 & 0.1237 & 0.7642 \\
& \multicolumn{7}{@{}l}{\emph{Calibration family}}\\
& \quad uncalibrated & 0.5274 & 0.1668 & 0.5710 & 0.7926 & 0.1237 & 0.7642 \\
& \quad global scaling 0.90 & 0.5188 & 0.1653 & 0.5645 & 0.8658 & 0.1065 & 0.9384 \\
& \quad gated 0.90 / 1.00 (adopted) & 0.5238 & 0.1657 & 0.5704 & 0.7952 & 0.1099 & 0.7716 \\
\bottomrule
\multicolumn{8}{@{}l}{\footnotesize $^{\dagger}$In-house ROI structural score; not the challenge ROI SSIM implementation.}
\end{tabular}
\end{table*}

\section{Discussion}

\textbf{Why a modular composition.} We arrived at the modular composition because no
single model family we evaluated provided the desired balance across global fidelity,
perceptual quality, and lesion-focused internal metrics. Each component is individually
unbalanced by design; what produced balance in our setting was a spatial prior
computable at test time together with a fusion rule that treats enhancement and texture
differently. The oracle analysis indicates how far this could be pushed: a perfect
lesion region would have been worth 8.22\% MSE in our internal evaluation, and since a
\emph{dilated} perfect region recovers only 1.47\%, the precision of the predicted
region---rather than the gate machinery around it---appears to be the limiting factor.
Improving the spatial prior therefore remains a substantial source of remaining
headroom, even though the localization improvements actually available to us had begun
to flatten.

\textbf{Design decisions driven by negative results.} Three internal negative results
changed the design more than any successful variant did.

\emph{Perceptual post-processing was restricted to the background.} Output sharpening
improved official LPIPS by 19.9\% on the fixed five-case debug cohort while degrading
the tumour-versus-non-tumour AUROC by 9.5\%. Perceptual gains that are not spatially
restricted are paid for in the lesion, where the downstream evaluators read. We
therefore admitted generative texture only through the soft background gate.

\emph{A single stronger network did not replace the ensemble.} A NAFNet-based
model~\cite{ref:nafnet} reached LPIPS 0.145 on our internal harness---better than both
the earlier ensemble and the submitted method---with near-perfect background
reconstruction, but an internal tumour-region structural score of 0.442, roughly 22\%
below the ensemble; a full-resolution retraining reached 0.489, short of our
pre-declared acceptance threshold of 0.50. In our internal evaluation, improving global
fidelity and perceptual quality did not preserve ROI fidelity, so we retained the
ensemble. A separate lesson came from the same work: the first run was evaluated in a
fixed $512\times512$ centre-cropped canvas, where MSE appeared to be 0.219 but was
0.545 at native resolution. All evaluation in this paper is at native resolution for
that reason.

\emph{Inter-slice context was moved into training rather than inference.} A 2.5D
teacher using adjacent slices improved tumour enhancement substantially
(tumour-under-enhancement reduced by 75.1\%) at substantial cost to global fidelity and
background behaviour (MSE $+72.8\%$, background over-enhancement $+28.8\%$). Direct
2.5D inference is also inadmissible under the single-slice contract, and the distilled
single-slice student was worse than the base network standalone. We therefore used
inter-slice information only as training-time augmentation
(Section~\ref{sec:data-masks}), where it did help.

\textbf{Challenge-engineering observations.} Two operational points generalise beyond
this task. Component-wise train/inference normalisation parity must be tested
explicitly in multi-component pipelines; our generator and regression branches lived in
different intensity spaces, and only a tensor-level parity check caught it before
packaging. And our internal proxies ranked two final candidates in the opposite order
to the official metrics: under a one-shot test budget, a small fixed-cohort official
measurement was worth more than any amount of internal comparison.

\textbf{Limitations.} Our internal validation set was reused for essentially every
selection decision and is a development set rather than a holdout, so internal values
are optimistic by an unquantified margin. The submitted method was never scored in the
challenge validation phase; its only pre-test official signal came from a five-case
debug cohort, which is too small to support inference. All ablations in this paper are
internal, so we cannot attribute any share of the official outcome to an individual
component, and we make no such claim. Some internal ablations used a different
lesion-probability aggregation rule and were therefore not bit-identical to the
submitted container; within a single harness that difference was worth approximately
2\% internal MSE, in the direction that the deployed mean-logit configuration is the
weaker of the two. Training data is US multi-centre while the hidden test cohort was
acquired at two external institutions in the Netherlands and Argentina at two fixed
matrix sizes, so shifts in acquisition settings and population characteristics apply. The
classification and segmentation metrics are properties of organizer-fixed evaluators
applied to our synthesised image and do not measure our own segmentation or
classification ability. The calibration constants are a global prior fitted on 120
development cases and are not validated per acquisition domain. The task is
single-slice and single-timepoint, and no reader study or clinical endpoint was
assessed.

\textbf{Future work.} The oracle result suggests a concrete next step: the precision of
the predicted lesion region is an important direction for further improvement, rather
than the gate machinery or further Dice gains of the kind that had already flattened.
Extending the gated composition to full volumes, and validating the calibration per
acquisition domain, are natural follow-ups.

\section{Conclusion}

We developed a lesion-gated modular solution for virtual contrast enhancement in breast
MRI that balanced complementary failure modes of regression and perceptual synthesis,
and ranked sixth in the MAMA-SYNTH Challenge. A lesion probability map estimated from
the pre-contrast slice alone---and therefore computable at test time---spatially
coordinates a tumour-focused regression pathway and a background-focused Pix2PixHD
synthesis pathway, closed by a region-dependent calibration of the predicted
enhancement. We adopted this modular composition because no single model family we
evaluated provided the desired balance across global fidelity, perceptual quality, and
lesion-focused internal metrics. The method was evaluated on the hidden external
300-case test cohort. Our internal ablation indicates that both localization quality
and gating strategy affected synthesis performance, that localization gains showed
diminishing returns at higher Dice, and that a ground-truth-mask oracle leaves
substantial remaining headroom associated with the precision of the predicted lesion
region.

\section*{Code Availability}
The implementation used for the MAMA-SYNTH Challenge is currently maintained in a
private repository. Methodological details and the trained inference pipeline are
described in this paper; code release will be considered following the post-challenge
publication process.

\section*{Funding}
This work received no external funding.

\section*{Competing Interests}
The author declares no competing interests.


\end{document}